\documentstyle[psfig]{mn}

\title[Optical and X-ray observations of OY Car]{An eclipse of the X-ray flux from the dwarf nova OY Carinae in quiescence}

\author[G.W. Pratt et al.]
	{Gabriel W. Pratt,$^1$ B.J.M. Hassall,$^1$ T. Naylor$^2$ and Janet H. Wood$^2$ \\
	 $^1$Centre for Astrophysics, University of Central Lancashire, Preston PR1 2HE, U.K.\\
	 $^2$Department of Physics, Keele University, Keele, ST5 5BG, U.K. }
         
\begin{document}

\maketitle


\begin{abstract}

We present a phase-resolved {\it ROSAT\/} HRI X-ray light curve of the dwarf nova OY Car in quiescence. The X-ray flux is eclipsed at the same time as the optical eclipse of the primary, and the region of X -ray emission is comparable in size to the white dwarf. We use subsequent optical observations to update the orbital ephemeris of the system. 

\end{abstract}




\section{Introduction}

Many non-magnetic cataclysmic variables (CVs), binary stars in which mass transfer is occurring between a Roche lobe-filling secondary star and a white dwarf primary, have been detected in X-rays. Steady state theory suggests that half the accretion luminosity is released in the accretion disc of such a system, whilst the other half is released in the region of the boundary layer, where the material of the accretion disc is decelerated to match the surface velocity of the primary. In quiescence, the boundary layer is believed to be an optically thin gas emitting relatively hard ($\sim 2 - 20$ keV) X-rays. In outburst, the accretion rate rises significantly, and the X-ray region becomes optically thick, now emitting softer ($\sim 0.1 - 1.0$ keV) X-rays \cite{pringsav79}. Provided the white dwarf is not spinning rapidly, theory predicts that the boundary layer is the dominant source of the X-ray emission in non-magnetic CVs \cite{pring77}. 

The dwarf novae are a subset of the non-magnetic CVs which are observed to undergo outbursts of $\sim 2 - 5$ mag on timescales of weeks to years. SU UMa dwarf novae also undergo additional infrequent superoutbursts, brighter by $\sim 1$ mag and lasting $\sim5$ times longer than normal outbursts. See Warner (1995) for a comprehensive review.

\begin{figure*}
\centerline{\psfig{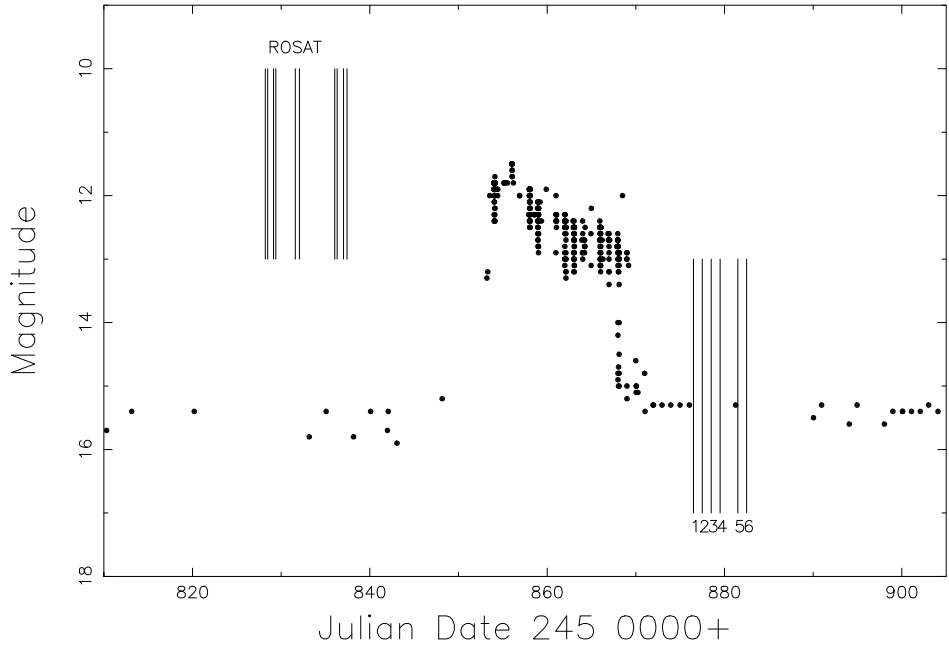}}
\label{fig1}
\caption{The light curve of OY Car, covering the period of both the {\it ROSAT} and optical observations. Circles show observations as supplied by the VSS RAS NZ. It can be seen that OY Car actually underwent a superoutburst that reached maximum on 1998 Feb 10. but had returned to quiescence by the time of our optical observations.}   
\end{figure*}

OY Carinae is among the the best-studied of the SU UMa dwarf novae, mainly because the favourable inclination of the system allows eclipses of the white dwarf, bright spot and accretion disc. In principle, the simplest observational test to determine the source of the X-rays is to observe systems in which the inclination is such that these full eclipses occur. Then if the X-ray flux is eclipsed, as it should be with a simple equatorial boundary layer around the white dwarf, the location of the X-ray source can be pinpointed to a relatively high degree of accuracy by comparing the timings of the X-ray eclipse with optical light curves. 


There are several high inclination dwarf novae that can be tested in this way. The {\it ROSAT} observation of HT Cas in a low state by Wood et al. \shortcite{woodetal95} showed that the X-ray source was indeed eclipsed, and that as expected, the X-ray eclipse is coincident with the optical eclipse of the white dwarf. Follow up observations of the system in quiescence using both {\it ASCA} and {\it ROSAT\/} \cite{mukaietal97} again showed an eclipse. The high time resolution of the {\it ASCA} data even allowed Mukai et al. (1997) to establish an approximate upper boundary layer dimension of $\sim 1.15 R_{\rm wd}$. An X-ray eclipse has also been found in {\it ROSAT} observations of the quiescent Z Cha \cite{vant97}.

In contrast, observations of the high accretion rate cases show that X-ray eclipses do not occur. This was the case with the {\it ROSAT\/} observations of the novalike variable UX UMa (Wood, Naylor \& Marsh 1995). Observations of dwarf novae in outburst or superoutburst are difficult due to their inherent unpredictability; of the high inclination systems, only OY Car has been observed in X-ray in superoutburst, and on both occasions there was no evidence for an eclipse of the X-ray flux (Naylor et al. 1988; Pratt et al. 1999). It appears that in these cases the X-ray emitting boundary layer is obscured, and the X-rays we see have an origin in a more extended, coronal source. The present challenge is to disentangle the effects of the disc, which becomes considerably thicker in the outburst state (e.g. Robinson et al. 1995, Pratt et al. 1999), and the increased local absorption due to the enhanced accretion, likely to be material in the upper atmosphere of the disc. 

Indeed, in the case of OY Car, it is the question of the absorption to the X-ray source that is particularly intriguing. {\it HST} observations in quiescence \cite{hornetal94} have suggested the existence of a significant local absorption due to disc material veiling the spectrum of the white dwarf, which Horne et al. (1994) dubbed the `iron curtain' after Shore (1992). Horne et al. (1994) modelled this material taking all relevant opacity sources into account, not just the Fe, but found it convenient to refer to the veiling material as the `iron curtain' due to the prevalence of blended Fe{\scriptsize II} features in the spectrum. 

\begin{figure}
\centerline{\psfig{figure=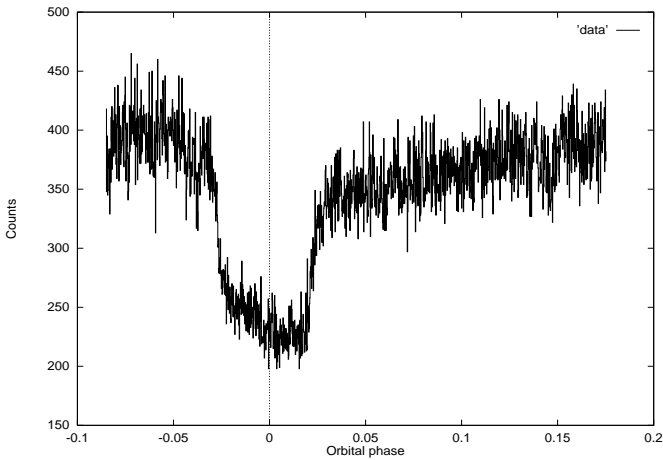,height=6cm,width=9cm,angle=0}}
\label{fig2}
\caption{Typical optical light curve (1998 March 4), phased with the ephemeris derived in Section~\ref{sec:neweph}.}   
\end{figure}

The aim of these observations is thus twofold: to determine if the `iron curtain' in OY Car in quiescence has sufficient column density to extinguish the X-ray flux in the bandpass of the {\it ROSAT} detectors ($0.1 - 2.5$ keV), and if not, to examine whether or not OY Car, like HT Cas and Z Cha, exhibits an X-ray eclipse.

In this paper, we present phase-resolved {\it ROSAT} X-ray observations of OY Car in quiescence. Section~\ref{sec:obsdata} describes the optical and X-ray observations and data reduction, in Section~\ref{sec:neweph} we derive an updated orbital ephemeris for the system, Section~\ref{sec:xecl} discusses the X-ray data, in Section~\ref{sec:discuss} we discuss our results, and in Section~\ref{sec:concs} we present our summary and conclusions.

\begin{figure*}
\centerline{\psfig{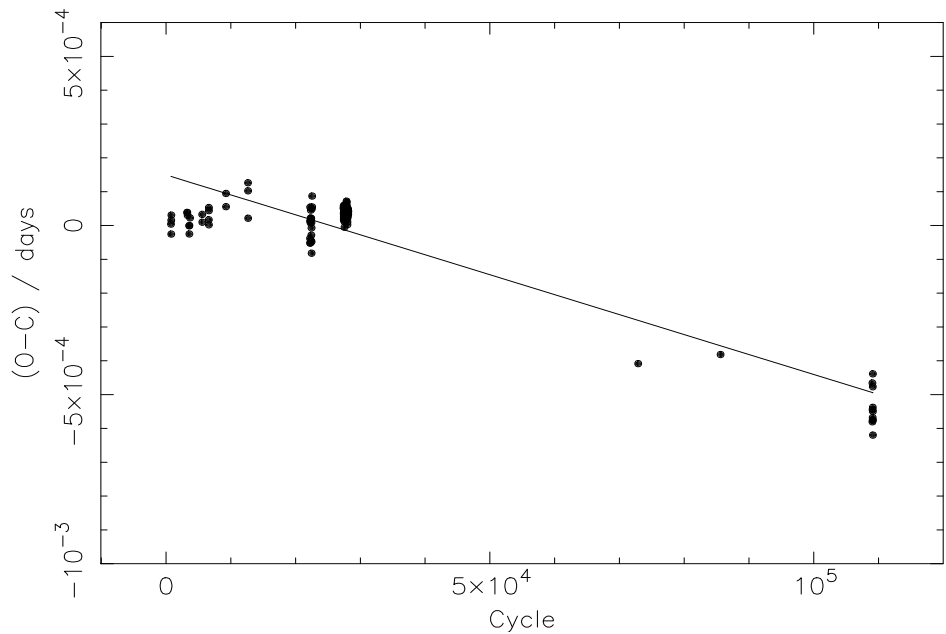}}
\label{fig3}
\caption{The {\it O-C\/} diagram of OY Car. The residuals are calculated with respect to the linear ephemeris of Wood et al. (1989). The solid line represents the fit to the residuals corresponding to a constant period with a phase offset.}   
\end{figure*}


\section{Observations and data reduction}

\label{sec:obsdata}

\subsection{Optical observations}

Previous optical and {\it HST\/} UV observations of OY Car obtained in 1992 and 1994 (Horne et al. 1994; Pratt et al. 1999) show that the ephemeris of OY Car as calculated in Wood et al. \shortcite{woodetal89} had drifted by $\sim 40$s at this time. Further observations of OY Car in quiescence were needed to update the ephemeris.

High-speed photometry of OY Car was obtained with the 1.0m telescope at the South African Astronomical Observatory (SAAO), between 1998 March 3 and 1998 March 9. Our data demonstrated that the system was in quiescence at the time. According to monitoring observations made by the Variable Star Section of the Royal Astronomical Society of New Zealand, the last superoutburst had reached maximum on 1998 Feb 10 and the following outburst/superoutburst had not occured at the time of writing (1998 Sept). Figure~1 shows the overall VSS RAS NZ light curve of OY Car covering the period of our optical and X-ray observations. The single-channel SAAO St. Andrews Photometer was used with a 16 arcsec aperture at 1s time resolution. For much of the duration of the observing run, conditions were non-photometric, but as  the primary aim of the run was the updating of the ephemeris, we continued observing through thin cloud. In most cases a $V$ filter was used, with occasional observations in difficult conditions made in white light. Sky measurements were taken every 15-20 min, at a position $\sim 1$ arcmin to the south of the source. A typical eclipse light curve is shown in Figure~2.


\subsection{X-ray observations}


{\tiny
\begin{table*}
\begin{minipage}{150mm}
\center
\caption{{\small Journal of X-ray observations.}}
\begin{tabular}{|c|c|c|c|r|r|}
\hline

\multicolumn{1}{|l}{{\it ROSAT\/}} & \multicolumn{1}{|c}{Date} & \multicolumn{1}{|c}{HJED start} &\multicolumn{1}{|c|}{Phase coverage} & \multicolumn{1}{c|}{Duration} & \multicolumn{1}{|c}{Source count rate}\footnote{This is the time-averaged count rate from the source after background subtraction. The errors quoted are Poissonian} \\ 

\multicolumn{1}{|l}{Observation ID } & \multicolumn{1}{|l}{ } & \multicolumn{1}{|c}{ } & \multicolumn{1}{|c|}{} & \multicolumn{1}{c|}{(seconds)} & \multicolumn{1}{|c}{(counts s$^{-1} \times 10^{-2}$)} \\ 

          &             & (2450800+)  &               &      &               \\ 
 rh300600 & 1998 Jan 14 & 28.216 & 0.875 - 0.146 & 1435 & $2.1 \pm 0.4$ \\
 rh300601 & 1998 Jan 14 & 28.476 & 0.990 - 0.004 & 83   & $2.4 \pm 0.9$ \\
 .        & .           & 28.481 & 0.007 - 0.348 & 1456 & $1.8 \pm 0.4$ \\
 .        & .           & 28.543 & 0.006 - 0.404 & 1856 & $1.9 \pm 0.3$ \\
 .        & .           & 28.620 & 0.282 - 0.452 & 928  & $1.3 \pm 0.4$ \\
 .        & .           & 28.666 & 0.009 - 0.165 & 848  & $1.1 \pm 0.4$ \\
 rh300602 & 1998 Jan 15 & 29.145 & 0.588 - 0.855 & 1440 & $1.3 \pm 0.3$ \\
 .        & .           & 29.212 & 0.653 - 0.902 & 1280 & $0.7 \pm 0.2$ \\
 rh300603 & 1998 Jan 15 & 29.402 & 0.667 - 0.006 & 1920 & $1.0 \pm 0.2$ \\
 .        & .           & 29.470 & 0.737 - 0.792 & 288  & $0.0 \pm 0.2$ \\
 .        & .           & 29.475 & 0.819 - 0.103 & 1488 & $1.9 \pm 0.4$ \\
 .        & .           & 29.538 & 0.813 - 0.160 & 1872 & $2.0 \pm 0.3$ \\
 rh300604 & 1998 Jan 18 & 31.598 & 0.453 - 0.477 & 96   & $3.4 \pm 1.9$ \\
 .        & .           & 31.605 & 0.565 - 0.717 & 832  & $1.4 \pm 0.4$ \\
 .        & .           & 31.652 & 0.316 - 0.477 & 864  & $2.9 \pm 0.6$ \\
 .        & .           & 31.673 & 0.636 - 0.770 & 736  & $1.7 \pm 0.5$ \\
 .        & .           & 31.714 & 0.296 - 0.550 & 1392 & $1.8 \pm 0.4$ \\
 .        & .           & 31.741 & 0.715 - 0.818 & 560  & $1.0 \pm 0.4$ \\
 rh300605 & 1998 Jan 18 & 32.061 & 0.794 - 0.007 & 1520 & $1.3 \pm 0.3$ \\
 .        & .           & 32.131 & 0.891 - 0.120 & 1248 & $0.6 \pm 0.2$ \\
 .        & .           & 32.195 & 0.907 - 0.176 & 1408 & $0.3 \pm 0.4$ \\
 .        & .           & 32.254 & 0.840 - 0.224 & 1824 & $1.2 \pm 0.3$ \\
 rh300606 & 1998 Jan 19 & 33.055 & 0.543 - 0.831 & 1574 & $1.1 \pm 0.3$ \\
 .        & .           & 33.122 & 0.612 - 0.877 & 1408 & $1.4 \pm 0.3$ \\
 .        & .           & 33.172 & 0.385 - 0.932 & 2895 & $2.3 \pm 0.3$ \\
 rh300607 & 1998 Jan 20 & 33.516 & 0.845 - 0.181 & 1834 & $2.1 \pm 0.3$ \\
 .        & .           & 33.569 & 0.688 - 0.972 & 1378 & $0.9 \pm 0.2$ \\
 .        & .           & 33.594 & 0.008 - 0.226 &  803 & $2.2 \pm 0.5$ \\
 .        & .           & 34.121 & 0.420 - 0.632 & 1126 & $1.6 \pm 0.4$ \\
 .        & .           & 34.182 & 0.399 - 0.688 & 1522 & $1.3 \pm 0.3$ \\
 rh300608 & 1998 Jan 22 & 36.036 & 0.771 - 0.009 & 1680 & $1.0 \pm 0.2$ \\
 .        & .           & 36.102 & 0.815 - 0.144 & 1664 & $2.5 \pm 0.4$ \\
 rh300609 & 1998 Jan 22 & 36.316 & 0.207 - 0.301 & 512  & $0.8 \pm 0.4$ \\
 .        & .           & 36.364 & 0.964 - 0.345 & 2000 & $2.0 \pm 0.3$ \\
 .        & .           & 36.431 & 0.002 - 0.390 & 1968 & $1.8 \pm 0.3$ \\
 rh300610 & 1998 Jan 23 & 37.030 & 0.514 - 0.857 & 1840 & $1.3 \pm 0.3$ \\
 .        & .           & 37.097 & 0.576 - 0.908 & 1696 & $1.2 \pm 0.3$ \\
 rh300611 & 1998 Jan 23 & 37.426 & 0.778 - 0.159 & 2064 & $1.5 \pm 0.3$ \\

\hline
\end{tabular}
\label{tab:xray_obs}

\end{minipage}
\end{table*}
}


\begin{table}
\center
\caption{{\small Journal of optical observations.}}
\begin{tabular}{|l|c|l|l|l}
\hline

\multicolumn{1}{|l}{Run} & \multicolumn{1}{|l}{Date} & \multicolumn{1}{|c}{UT start} & \multicolumn{1}{|c}{Cycle} & \multicolumn{1}{|c}{Filter}\\

\multicolumn{1}{|c}{ } & \multicolumn{1}{|l}{} & \multicolumn{1}{|l}{} & \multicolumn{1}{|l}{} \\

 0007 & 1998 March 3 & 22:57:30 & 109036 & $V$ \\ 
 0009 & 1998 March 4 & 01:22:43 & 109045 & $V$ \\
 1008 & 1998 March 5 & 01:46:17 & 109061 & $V$ \\
 2001 & 1998 March 6 & 00:33:42 & 109076 & $V$ \\
 3009 & .            & 18:41:45 & 109088 & $V$ \\
 3017 & .            & 23:10:31 & 109091 & $V$ \\
 5015 & 1998 March 8 & 20:42:03 & 109121 & $V$ \\
 5019 & .            & 21:55:27 & 109122 & $V$ \\
 5024 & .            & 23:35:29 & 109123 & $V$ \\
 5028 & 1998 March 9 & 01:09:48 & 109124 & $V$ \\
 5030 & .            & 02:07:23 & 109125 & $V$ \\
 6013 & .            & 18:53:30 & 109136 & $W$ \\

\hline
\end{tabular}
\label{tab:opt_obs}
\end{table}

OY Car was observed 12 times by the {\it ROSAT\/} High Resolution Imager (HRI) for a total of 51.5 kiloseconds between 1998 January 14 and 1998 January 23. Information about the individual observations is presented in Table~\ref{tab:xray_obs}, the journal of X-ray observations. 

The orbital period of OY Car, at $\sim 91$ min, is very similar to the orbital period of {\it ROSAT}. As a consequence of this, the orbital phase of OY Car covered by {\it ROSAT} precesses by $\sim 5$ min per satellite orbit, leading to a very similar phase coverage for a set of contiguous observations. Additional difficulties in obtaining the desired orbital phase coverage include Earth occultations and the passage of the satellite through the South Atlantic Anomaly. Our main aim during the course of these observations was sufficient coverage of the eclipse phase (i.e., $-0.25 \la \phi \la 0.25$), with additional coverage of the full orbital phase being of secondary importance. 

The data were reduced using {\scriptsize ASTERIX} software. For each observation, source counts were extracted from a circular region of radius 50 arcsec. For the determination of the background, we used an annular region centred on the source, with inner and outer radii of 100 arcsec and 300 arcsec, respectively. Source and background data were extracted in 1s bins and the data were then background subtracted. The resulting background subtracted total flux light curves were then corrected for vignetting, dead time, and scattering. We then converted all times to HJED. 

In addition to extracting the source counts in this way, the extracted background data were scaled to the dimensions of the source extraction circle to create a background light curve.


\section{The orbital ephemeris}

\label{sec:neweph}

\subsection{Optical eclipse timing measurements}

In all, we observed 12 useful optical eclipses, details of which can be found in Table~\ref{tab:opt_obs}. As we were only interested in eclipse timings, the data reduction process was simplified considerably. For each eclipse light curve, all times were converted to Heliocentric Julian Ephemeris Date (HJED). The times of mid-eclipse were then calculated automatically using the method described in Wood, Irwin \& Pringle \shortcite{woodetal85}. The light curves were first smoothed by passing them through a median filter of width 10s, roughly a quarter of the duration of the ingress of the white dwarf. Each light curve was then differentiated numerically, and the differential passed through a box car filter of width 42s. This filter width was chosen as it is the approximate duration of the ingress of the white dwarf (c.f., Wood et al. 1985; Wood et al. 1989). This filtered derivative was then examined to detect the largest negative and positive points, which correspond to the mid-points of ingress and egress respectively. Mid-eclipse was then calculated as being the point halfway between the ingress and egress mid-points, i.e., $\phi_{0} = (\phi_{\rm wi} + \phi_{\rm we})/2$. The HJED times of mid-eclipse are listed in Table~\ref{tab:opt_ecl}.


\begin{table}
\begin{minipage}{80mm}
\center
\caption{{\small Timings of optical mid-eclipse.}}
\begin{tabular}{|l|l|c|}
\hline

\multicolumn{1}{|l}{Cycle} & \multicolumn{1}{|c}{HJED} & \multicolumn{1}{|c}{$O-C$}\\

\multicolumn{1}{|c}{ } & \multicolumn{1}{|l}{(2400000+)} & \multicolumn{1}{|r}{($10^{-4}$ day)}\\

        &              &       \\
 85618  & 49397.84069\footnote{Eclipse timing from a near quiescent light curve in Pratt et al. (1999).} & -3.8\\
        &              &       \\
 109036 & 50876.51124  & -5.8  \\
 109045 & 50876.57449  & -4.7  \\
 109061 & 50877.58431  & -5.8  \\
 109076 & 50878.53116  & -5.4  \\
 109088 & 50879.28858  & -5.8  \\
 109091 & 50879.47795  & -5.7  \\
 109121 & 50881.37157  & -5.8  \\
 109122 & 50881.43479  & -4.8  \\
 109123 & 50881.49785  & -5.4  \\
 109124 & 50881.56107  & -4.4  \\
 109125 & 50881.62408  & -5.5  \\
 109136 & 50882.31834  & -6.2  \\

\hline
\end{tabular}
\label{tab:opt_ecl}

\end{minipage}
\end{table}

\subsection{Method}

In order to update the orbital ephemeris of OY Car, we used published eclipse timings ranging back to 1979 in addition to those timings presented in Table~\ref{tab:opt_ecl}.

The (O-C) values were computed with respect to the first linear ephemeris of Wood et al. (1989; this paper also contains a two-part linear ephemeris), and these are plotted in Figure~3. Eclipse timings from Vogt et al. \shortcite{vogtetal81}, Schoembs \& Hartmann \shortcite{schart83}, Cook \shortcite{cook85}, Schoembs, Dreier \& Barwig \shortcite{schdribar87}, Wood et al. \shortcite{woodetal89}, Horne et al. \shortcite{hornetal94}, and Pratt et al. \shortcite{prattetal99a} are plotted in addition to the eclipse timings presented here. Where needed, the published eclipse timings were converted to HJED.

Figure~3 shows the result of a linear least squares fit represented by the following updated ephemeris (uncertainties quoted in parentheses):

\[
{\rm HJED} = 2 443 993.553 958(6) + 0.0631209180(2){\rm E}
\]

This linear ephemeris fit to observations over a baseline of 19 years suggests that the two-part ephemeris calculated in Wood et al. \shortcite{woodetal89} is unnecessary. A quadratic ephemeris offered no significant improvement to the fit.

\begin{figure}
\centerline{\psfig{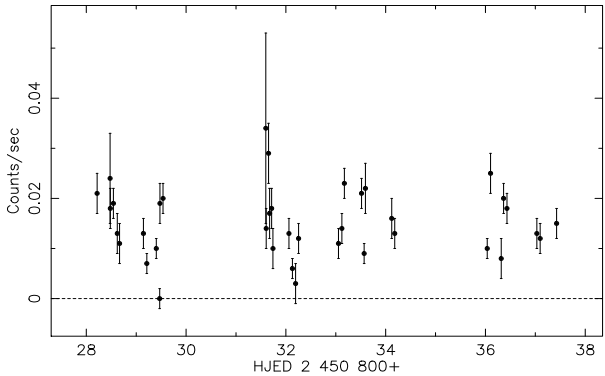}}
\label{fig4}
\caption{The behaviour of the X-ray flux over the duration of the {\it ROSAT} observations. The panel shows the data averages with $1\sigma$ error bars. Eclipses have not been removed.}   
\end{figure}

We note that the data used by Wood et al. \shortcite{woodetal89} describes a totally different ephemeris. Unfortunately, the sampling of the data presented here precludes the establishment of a model for the evolution of the ephemeris of OY Car, a common problem with this type of observation. It is known that the ephemerides of both HT Cas and IP Peg have been observed to undergo apparently uncorrelated departures from their assumed linear paths. It is not known what causes these departures, although changes in the radius of the secondary could conceivably provide a mechanism (e.g. Applegate 1992). 

This updated ephemeris is used throughout the remainder of this paper for the purposes of phasing the X-ray observations.



\section{The X-ray eclipse}

\label{sec:xecl}

Figure~4 shows that the X-ray flux is variable over the duration of the observations. The data cover 10 full eclipses and 9 partial eclipses. The mean HRI count rate over the duration of the observations is $(1.60 \pm 0.06) \times 10^{-2}$ counts per second.


\begin{figure}
\centerline{\psfig{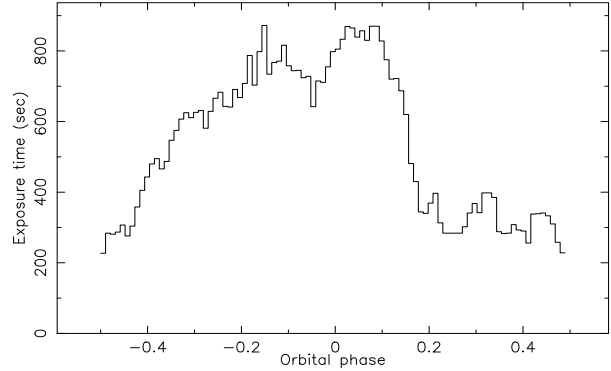}}
\label{fig6}
\caption{The efective exposure times of each of the 96 phase bins used for the superoutburst light curve, in seconds.}   
\end{figure}

We obtained the mean quiescent X-ray light curve by folding all the available data into 96 phase bins on the orbital period of OY Car. The total exposure time of each of the 96 phase bins is shown in Figure~5. The duration of the optical white dwarf eclipse, defined as the difference between the second and third contact phases of the white dwarf, is 231 s (Wood et al. 1989). Thus with 96 phase bins, four bins cover the eclipse and have a total duration of 227 s. The bins were chosen such that the start of one bin coincided with orbital phase zero.

In Figure~6, we show the quiescent X-ray light curve of OY Car, binned into 96 phase bins, phased on the ephemeris presented in this paper with a correction of -8s subtracted from the constant term. With the -8s correction, the $\sigma-$values of all four eclipse bins are at a simultaneous low, implying that this is the best ephemeris for the X-ray data. The spread of values in the optical $O-C$ determinations in Section~\ref{sec:neweph} is $\sim 2.4 \times 10^{-4}$ of a day, or 16s, so the discrepancy in the ephemeris of the X-ray data is well within the error of the optical observations.



\begin{figure*}
\centerline{\psfig{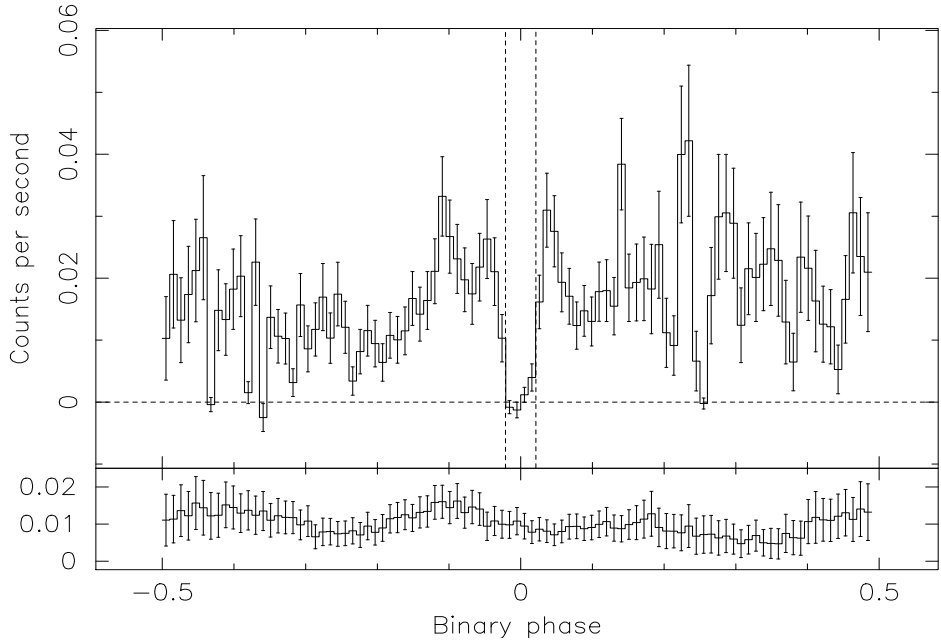}}
\label{fig5}
\caption{The {\it ROSAT\/} HRI quiescent X-ray light curve, folded on the orbital period of OY Car in 96 bins, is shown in the top panel. The bottom panel shows the corresponding folded background light curve, scaled to the dimensions of the source extraction circle. The dotted lines represent the times of the optical white dwarf eclipse (Wood et al. 1989). The error bars are $1\sigma$ Poissonian. The data have been phased using the ephemeris derived in this paper with an adjustment of -8s. See text for explanation.}   
\end{figure*}

To investigate the significance of the eclipse, the data were binned into 24 bins such that the duration of one bin was marginally smaller than the duration of the optical white dwarf eclipse. Now the out of eclipse flux was $(1.70 \pm 0.06) \times 10^{-2}$ counts s$^{-1}$, and the flux in eclipse was $(0.14 \pm 0.06) \times 10^{-2}$ counts s$^{-1}$. The eclipse is thus significant at the $2.3\sigma$ level. The fact that the flux drops at exactly phase zero adds significance to the detection.

We used two methods to estimate the duration of the X-ray eclipse. First, we divided the data into 96 phase bins and shifted the bins backwards and forwards until in one of the four eclipse bins, the count rate deviated from zero by $2.5\sigma$ or greater (c.f. van Teeseling 1997). This allows us to estimate an upper limit to the width of the X-ray eclipse of $\sim 271$s, which is only 40s longer than the optical eclipse. Secondly, assuming that the eclipse is symmetric,  we examined the symmetrized light curve folded at 20s time resolution (c.f. Wood et al. 1995a). The full width zero depth of the eclipse in this case is then FWZD = 259s, or only 28s longer than the optical eclipse of the white dwarf. 

Finally, we note the apparent similarities between the average X-ray light curve presented here and that obtained for Z Cha by van Teeseling (1997), in that both light curves exhibit a dip near phase 0.7 --- 0.8. While this is suggestive of flux variability with phase, the low signal to noise precludes a quantitative analysis of these features, even in the average light curve.


\section{Discussion}

\label{sec:discuss}

Given that X-ray eclipses have been detected in HT Cas and Z Cha in quiescence, it is perhaps not surprising to have found the same for OY Car. But it is OY Car which has recently been the focus of X-ray and UV investigations which have raised some interesting questions concerning the local absorption of the system. The {\it HST\/} UV observation by Horne et al. \shortcite{hornetal94} showed an `iron curtain' of veiling material superimposed on the spectrum of the white dwarf. Horne et al. \shortcite{hornetal94} cannot pinpoint the exact position of this gas, although placing the material at the outer disc allows the more plausible explanation that the disturbances that produce the material are driven by the gas stream. Independent of the location of the gas, Horne et al. (1994) fit a column density of $n_{H} \simeq 10^{22}$ cm$^{-2}$. However, at this column density, the material in front of the white dwarf should absorb all the X-rays (see e.g. Naylor \& la Dous 1997).

In this paper we show that there is a clear and unambiguous (although not necessarily total) eclipse of the X-rays, demonstrating the column density of the intervening material is not sufficient to extinguish the flux in the {\it ROSAT\/} passband. 

Variability could be the key: it may be that the `iron curtain' is not a permanent feature and that the Horne et al. \shortcite{hornetal94} observation happened to catch the system when the column density of the intervening material was higher than usual. One obvious cause of a higher local absorption than usual is a recent outburst or superoutburst: however the Horne et al. (1994) observation was obtained in the middle of a quiescent period, strongly suggesting that the source of the veiling gas is not connected to the outbursts. 

An alternative explanation is variability on the orbital timescale. Because of the nature of the existing observations, we do not know how the absorption changes with orbital phase: it is quite possible that the local absorption varies azimuthally around the disc. This can be seen in the {\it IUE} spectra of Z Cha, in which it can be seen that the `iron curtain' is present in some spectra but not in others (Wade, Cheng \& Hubeny 1994)

It does seem clear that the absorption is intimately connected to the accretion of material, and as this is variable, then so is the strength of the absorption. Three separate X-ray observations of HT Cas have yielded three widely differing values of $n_{H}$: $1.8 \times 10^{20}$ cm$^{-2}$ in the low state \cite{woodetal95}, but between $6\times 10^{20}$ and $3 \times 10^{21}$ cm$^{-2}$ in quiescence (Patterson \& Raymond 1985; Mukai et al. 1997).

Finally, our observation is similar to the previous detections of X-ray eclipses in dwarf novae in that the inferred size of the X-ray emitting region is comparable to the size of the white dwarf, and comes from a source close to the primary. We were, however, unable to establish a reliable lower limit for the duration of the X-ray eclipse. We now know that the X-rays in quiescence are definitely emitted from a region in the neighbourhood of the white dwarf, but we cannot be any more precise than that due to data constraints. 


\section{Summary and conclusions}

\label{sec:concs}

We have shown that there is an eclipse of the X-ray flux from OY Car in the quiescent state, and that the X-rays are emitted from a region very close to the white dwarf.

From observations of suitable systems, we now know that in general eclipses of the X-ray flux occur where the accretion rate is low, and do not occur where the accretion rate is high. We know that there are discrepancies between the X-ray and UV measurements of the local absorption in OY Car, and that the strength of the absorption appears to be variable and intimately connected to the accretion rate in the case of HT Cas. 

For OY Car we now need higher quality X-ray light curves that will enable us to pinpoint the exact source of the X-ray emission. Data with higher time and energy resolution would be desirable, such as is currently available from {\it ASCA}. Simultaneous orbital phase dependent observations of these systems in the UV, such as with the {\it HST}, will enable the investigation of the variability of the local absorption with orbital phase in both UV and X-rays. Looking further ahead, {\it XMM} will represent the next generation of X-ray satellites, and will be ideal for the study of these intriguing systems due to the higher throughput and especially the simultaneous optical monitoring. 


\section*{acknowledgements}

We thank Jakob Engelhauser for his help with the difficult scheduling of the {\it ROSAT} observations. We acknowledge the data analysis facilities provided by the Starlink Project, which is run by CCLRC on behalf of PPARC. GWP is in receipt of a University of Central Lancashire research studentship.



\end{document}